\documentclass[12pt]{article}

\usepackage{graphicx,hyperref}
\usepackage{amsmath}
\usepackage{amssymb}
\usepackage{amsthm}
\usepackage{setspace}
\usepackage{natbib}
\usepackage{marginnote}

\newcommand{\Polya}{P\'{o}lya}
\newcommand{\PG}{\text{PG}}

\newcommand{\NB}{\text{NB}}

\newcommand{\bbeta}{\boldsymbol{\beta}}
\newcommand{\oomega}{\boldsymbol{\omega}}
\newcommand{\yy}{\boldsymbol{y}}
\newcommand{\vx}{\boldsymbol{x}}
\newcommand{\betap}{{\boldsymbol{B}}}
\newcommand{\bW}{\boldsymbol{W}}
\newcommand{\bPhi}{\boldsymbol{\Phi}}

\newcommand{\bmu}{\boldsymbol{\mu}}

\newcommand{\vep}{\boldsymbol{\ep}}

\newcounter{parnum}
\newcommand{\npoint}{%
  \noindent\refstepcounter{parnum}%
  \makebox[0.5in][c]{\textbf{\arabic{parnum}.}} %
  \marginnote{\small\ttfamily\the\inputlineno}}
\renewcommand{\npoint}{}

\ifdefined \theorem
\else
\theoremstyle{definition}
\newtheorem{theorem}{Theorem}

\newtheorem{example}[theorem]{Example}

\fi

\newcommand{\mycomment}[1]{}

\newcommand{\ep}{\varepsilon}

\newcommand{\mcO}{\mathcal{O}}

\begin{document}

\renewcommand{\thefootnote}{\fnsymbol{footnote}}
\author{
  Jesse Windle\footnotemark[1],
  Carlos M. Carvalho\footnotemark[2],
  James G. Scott\footnotemark[2], and
  Liang Sun\footnotemark[2]%
}

\title{Efficient Data Augmentation in Dynamic Models for Binary and Count Data}

\maketitle

\footnotetext[1]{
 Duke University
 \href{mailto:jesse.windle@stat.duke.edu}{jesse.windle@stat.duke.edu}
}
\footnotetext[2]{
 The University of Texas at Austin
}
\renewcommand{\thefootnote}{\arabic{footnote}}


\begin{abstract}
Dynamic linear models with Gaussian observations and Gaussian states lead to
closed-form formulas for posterior simulation.  However, these closed-form
formulas break down when the response or state evolution ceases to be Gaussian.
Dynamic, generalized linear models exemplify a class of models for which this is
the case, and include, amongst other models, dynamic binomial logistic
regression and dynamic negative binomial regression.  Finding and appraising
posterior simulation techniques for these models is important since modeling
temporally correlated categories or counts is useful in a variety of
disciplines, including ecology, economics, epidemiology, medicine, and
neuroscience.  In this paper, we present one such technique, \Polya-Gamma data
augmentation, and compare it against two competing methods.  We find that the
\Polya-Gamma approach works well for dynamic logistic regression and for dynamic
negative binomial regression when the count sizes are small.  Supplementary
files are provided for replicating the benchmarks.

\noindent %
\end{abstract}


\doublespace

\section{Introduction}

\subsection{Bayesian inference for complex discrete data}

Bayesian inference for discrete-data models has long been recognized as a
challenging problem, due to the analytically inconvenient form of the likelihood
functions that arise.  This makes it much more difficult to fit discrete models
with rich stochastic structure.  This is in stark contrast to the case of
real-valued data, for which the Bayesian literature contains a tremendous
variety of highly structured models.  These tools allow us to handle data sets
that are not merely large, but also dense: varying in time or space, rich with
covariates, or layered with multi-level structure.  This level of sophistication
is made possible by the mathematical and computational simplicity of the
Gaussian likelihood, and the flexibility of the mixture-of-Gaussians class.

Given the widespread need for rich discrete-data models, much work has been done
on making them amenable to Bayesian inference.  The traditional approach is to
work directly with the discrete-data likelihood, whether via numerical
integration \citep{skene-wakefield-1990}, analytic approximations
\citep{carlin-1992,bradlow-etal-2002,gelman-etal-2008,forster-2011}, or the
Metropolis-Hastings algorithm \citep{dellaportas-forster-1999,
  dobra-tebaldi-west-2006}.  These methods are used in popular R packages that
implement MCMC for non-Gaussian models \citep{mcmcpack-2011}.  Unfortunately,
none lead to a fully automatic approach to posterior inference, as they require
either approximations (whose quality must be validated) or the choice of a
tuning constant (as in the Metropolis--Hastings sampler) that strongly affects
performance.  In practice, these difficulties limit users to simple models and
small data sets, especially in the case of non-i.i.d.~data.

This paper considers the particular situation of integer-valued outcomes
collected over time.  This kind of data may be found in, for example, ecology or
epidemiology when modeling populations of species or outbreaks of infections,
and in neuroscience when modeling brain activity as manifest in neural spike
trains.  Such data sets are archetypal of the wider pattern of computational
difficulty associated with discrete data sets: in addition to the non-Gaussian
likelihood, one must also account for the temporal correlation.
Autoregressive-like models are helpful in this regard, at the cost of making
posterior inference more difficult.

We directly address this problem, proposing an elegant and efficient data
augmentation technique for dynamic models with binomial likelihoods.  This
includes the logistic regression model and its variants (for binary and
categorical data), along with the negative-binomial model (for count data).  Our
approach involves a data-augmentation scheme based on the family of Polya-Gamma
distributions.  As we will show, this is the crucial step in allowing a wide
class of models to be handled using simple variations on established techniques
for Bayesian inference in linear Gaussian state-space models.


\subsection{Background}

\npoint The origin of dynamic logit and negative-binomial models can be traced
to their static ancestors, which, in turn, are successors to the linear model.
We provide a brief overview of this evolution and the inferential challenges
that arise along the way.

\npoint Generalized linear models (GLMs) 
parameterize the expectation and variance of the response in terms of a linear
combination of the predictors, $\psi_t = \vx_t' \bbeta$, and lead to tractable
models for counts, categories, and other non-real valued data
\citep{wedderburn-1974, mccullagh-nelder-1989}.  Strictly speaking, one need not
specify the distribution of the response, just its first two moments, but since
we are interested in Bayesian inference we will restrict our attention to
situations in which the likelihood $f(y_t | \psi_t)$ is specified.  When the
likelihood comes from the exponential family of distributions, one may easily
calculate the posterior mode and associated error estimates; however, the
corresponding posterior distribution is usually difficult to sample, requiring
Metropolis-Hastings sampling, Gibbs sampling via data augmentation, or some
other MCMC method.

\npoint Dynamic generalized linear models (DGLMs) permit an evolving
relationship between the response and the covariates via time-varying regression
coefficients, $\{\bbeta_t\}_{t=1}^T$, so that $\psi_t = \vx_t' \bbeta_t$ for
$t=1, \ldots, T$.  One must specify a prior for $\{\bbeta_t\}_{t=1}^T$, which
controls how quickly $\{\bbeta_t\}_{t=1}^T$ may change, to complete the model.
It is common to let $\{\bbeta_t\}_{t=1}^T$ be a Gaussian AR(1) process or a
Gaussian random walk.  In either case, one recovers a state-space model
characterized by the observation distribution $f(y_t | \psi_t)$ for the response
and the Markovian evolution distribution $g(\bbeta_t | \bbeta_{t-1})$ for the
hidden states.  Transitioning from the static to the dynamic model is a small
step conceptually but a big step practically; in particular, it becomes even
more difficult to generate posterior samples.

\npoint To clarify the challenge of simulating $(\{\bbeta_t\}_{t=1}^T |
\{y_t\}_{t=1}^T)$ for dynamic generalized linear models, it is helpful to
consider more and less tractable state-space models.  
\npoint State-space models with non-Gaussian, non-linear responses and
non-linear evolution equations are less tractable.  One may efficiently
approximate the filtered distributions of such models, $p(\bbeta_t |
\{y_s\}_{s=1}^t)$, using sequential methods, but the lack of structure makes
sampling from the posterior of $\{\bbeta_t\}_{t=1}^T$ difficult.  
\npoint In contrast, dynamic linear models (DLMs) with Gaussian states and
Gaussian observations are more tractable: one may sample the posterior
distribution of the states using the forward filter backward sample (FFBS)
algorithm \citep{carter-kohn-1994, fruhwirth-schnatter-1994}, a procedure that
is linear in the number of observations and thus quite efficient.  However, the
FFBS algorithm breaks down outside of these vary narrow assumptions.
\npoint Dynamic generalized linear models that have Gaussian, linear evolution
equations, but non-Gaussian, non-linear response distributions sit between these
extremes.  Within this class, the FFBS algorithm is not immediately applicable,
but the linear evolution equations give one hope of finding a clever way to
reintroduce it.  Many approaches follow this path, combining techniques used for
generalized linear models with the FFBS algorithm.  However, these amalgamations
are not always as effective as their constituent parts.  In particular,
Metropolis-Hastings based techniques for GLMs do not translate well to the
dynamic setting, as $\{\bbeta_t\}_{t=1}^T$ is of much higher dimension than
$\bbeta$.  
\npoint Data augmentation techniques for GLMs that lead to a Gaussian complete
conditional for $\bbeta$ provide a preferable approach since the corresponding
complete conditional for $\{\bbeta_t\}_{t=1}^T$ will coincide with the
likelihood of a DLM and hence the FFBS algorithm.  (See Section \ref{sec:pg} for
details.)  But such data augmentation tricks are difficult to find.  Recently,
\citet{polson-etal-2013} introduced a data augmentation trick for GLMs with
binomial likelihoods and that is what we exploit here.

\subsection{Previous efforts}
\label{sec:previous-efforts}


\npoint Approaches to posterior simulation date back to at least
\citet{west-etal-1985}, who employ a conjugate updating, backward sampling
(CUBS) strategy to generate approximate posterior draws.  Their strategy is to
use an approximation of the filtered distributions $(\bbeta_t |
\{y_s\}_{s=1}^t)$ when backwards sampling.  The CUBS method is also $\mcO(T)$ as
$T$ varies, but requires solving $T$ non-linear equations, which is time
consuming.

\npoint Metropolis-Hastings based approaches can take approximate draws and make
them exact by introducing an accept/reject step.  The challenge is to devise a
good approximation so that the probability of accepting a proposal is
reasonable.  As the dimension of the quantity one wants to sample increases,
this becomes more difficult.  A potential solution is to break the
high-dimensional quantity into pieces and sample the complete conditional of
these pieces sequentially, that is to do Metropolis-Hastings within Gibbs
sampling.  However, sampling the high-dimensional quantity in blocks tends to
increase the correlation between the successive samples within the Markov chain.
Thus, one must try to strike a balance between blocks that are too large,
leading to poor acceptance rates, and blocks that are too small, leading to
excessive autocorrelation.  One finds an extreme form of the blocking approach
in \citet{carlin-etal-1992}, prior to the advent of the FFBS.

\npoint Given this general strategy, one must still figure out how to pick a
good proposal distribution.  Since $f$ comes from the exponential family, it is
natural to use the Laplace approximation to arrive at a Gaussian proposal, as
this coincides with the iteratively reweighted least squares procedure for
generalized linear models \citep{gamerman-1997}.  One may follow a similar
strategy by first doing a change of coordinates to sample the disturbances
instead of the hidden states and then use a Laplace approximation
\citep{shephard-pitt-1997, gamerman-1998}.  Advocates of this strategy suggest
that the subsequent blocking possesses less intrinsic autocorrelation.
\citet{migon-etal-2013} show that the CUBS approximation of \citet{west-etal-1985}
is a good proposal, and based upon our own computational experiments we find
that, indeed, the approach of \citet{migon-etal-2013} is better than sampling
blocks of disturbances.  However, none of these Metropolis-Hastings proposals
are completely satisfactory.  In particular, \citet{migon-etal-2013} is still
time consuming, requiring $T$ non-linear solves for each MCMC iteration.

\npoint Data augmentation provides a preferable approach.  If one finds a data
augmentation scheme for the static problem, that is when $\psi_t = \vx_t'
\bbeta$, so that the complete conditional of $\bbeta$ is Gaussian, then the
corresponding complete conditional of $\{\bbeta_t\}_{t=1}^T$ for the dynamic
analog will coincide with a DLM and one may use the FFBS algorithm.
\npoint But finding such a scheme is difficult since it requires auxiliary
variables that (1) yield a Gaussian complete conditional for $\bbeta$ and (2)
can be sampled efficiently.  Examples that almost meet both criterions include
\citet{mcfadden-1974}, where (1) is not met, and \citet{holmes-held-2006}, where
(2) can be significantly improved.

\npoint Fr\"{u}hwirth-Schnatter and Fr\"{u}hwirth and their colleagues Fussl,
Held, Rue, and Wagner have developed fixes for these shortcomings that rely upon
approximating distributions using discrete mixtures of Gaussians.  Their work
has lead to data augmentation schemes that satisfy (1) and (2), in an
approximate though accurate sense, for binomial and multinomial logistic
regression \citep{fruhwirth-schnatter-fruhwirth-2007,
  fruhwirth-schnatter-fruhwirth-2010, fussl-etal-2013}, Poisson regression
\citep{fruhwirth-schnatter-wagner-2006, fruhwirth-schnatter-etal-2009}, and
negative binomial regression \citep{fruhwirth-schnatter-etal-2009}.  (In the
sequel, for dynamic binomial logistic regression, we will limit our comparison
to \citet{fussl-etal-2013} since it appears to be the best choice within
Fr\"{u}hwirth-Schnatter et al.'s discrete mixture cornucopia.)
\npoint While their methods work well, several rely upon precomputing large
tables of weights, means, and variances for the components of a collection of
mixtures that approximate an entire family of distributions.  Further all of the
discrete mixture of normal techniques make use of at least two layers of
auxiliary variables.  One would prefer to avoid many layers of latents since
this may inhibit traversing the posterior landscape and enlarges the memory
footprint when storing the latent states.

\npoint The situation is significantly harder once one abandons the structure we
have assumed to this point.  \citet{geweke-tanizaki-2001} have reviewed a
plethora of approaches within the more general setting of state-space models,
none of which work that well.  \citet{godsill-etal-2004} show how to sample
smoothed states, $(\{\bbeta_t\}_{t=1}^T | \{y_t\}_{t=1}^T)$, using particle
methods; however, this is relatively expensive in comparison to filtering
states.  Recently, \citet{geweke-etal-2013} leveraged the power of GPUs as a way
to significantly speed up sequential Monte Carlo, an interesting avenue not
considered herein.

\section{\Polya-Gamma data augmentation}
\label{sec:pg}

We consider DGLMs with binomial likelihoods: \( f(y_t | q_t) = c(y_t) q_t^{a_t}
(1-q_t)^{d_t}, \) where $a_t$ and $d_t$ may depend on $y_t$ but do not depend
upon $q_t$.  Logistic regression and negative binomial regression are two common
models that fit within this regime.  It is preferable to express the likelihood
$f$ in terms of the log odds, $\psi_t = \log \frac{q_t}{1-q_t}$, since this is
the scale on which we linearly model the covariates:
\[
f(y_t | \psi_t) = c(y_t) \, \frac{\exp(\psi_t)^{a_t}}{(1+\exp(\psi_t))^{b_t}},
\]
where $b_t = a_t + d_t$.  Given a collection of observations $\yy =
\{y_t\}_{t=1}^T$, the posterior distribution of the hidden states $\betap =
\{\bbeta_t\}_{t=1}^T$ is
\[
p(\betap | \yy) = c(\yy) \Big[ \prod_{t=1}^T
\frac{\exp({\psi_t})^{a_t}}{(1+\exp({\psi_t}))^{b_t}}
\Big] p(\betap),
\]
where $\psi_t = \vx_t' \bbeta_t$.  (We will use the generic function $p$ to
denote a probability density.)  Following \citet{polson-etal-2013}, one may
introduce a collection of independent \Polya-Gamma random variates $\oomega =
\{\omega_t\}_{t=1}^T$, $\omega_t \sim \PG(b_t, \psi_t)$ for $t=1, \ldots, T$, to
construct the joint distribution
\[
p(\betap, \oomega | \yy) = c(\yy) \Big[ \prod_{t=1}^T
\frac{\exp({\psi_t})^{a_t}}{(1+\exp({\psi_t}))^{b_t}} p(\omega_t | b_t, \psi_t)
\Big] p(\betap) \, .
\]
via the conditional structure $p(\betap, \oomega | \yy) = p(\oomega | \betap,
\yy) p(\betap | \yy)$.  The $\PG{}(b_t, \psi_t)$ density possesses the special
form
\[
p(\omega_t | b_t, \psi_t) = \cosh^{b_t}(\psi_t/2) \exp({- \omega_t \psi_t^2 / 2})
p(\omega_t),
\]
which is useful since the ratio
\begin{equation}
  \label{eqn:pg-cancellation}
  \cosh^{b_t}(\psi_t/2) / (1+\exp({\psi_t}))^{b_t}
  \propto
  \exp({-\psi_t b_t/2}) 
\end{equation}
so that, upon completing the square, the complete conditional of $\betap$ is
\[
p(\betap | \oomega, \yy) \propto \Big[ \prod_{t=1}^T \exp \Big(-\frac{\omega_t}{2}
\Big(\frac{\kappa_t}{\omega_t} - \psi_t\Big)^2 \Big) \Big] p(\betap), \; \psi_t
= \vx_t' \bbeta_t,
\]
where $\kappa_t = a_t - b_t / 2$.  A single term from the product above is
identical to the likelihood of a pseudo-data point $z_t = \kappa_t / \omega_t$
drawn from $z_t \sim N(\psi_t, 1/\omega_t)$.  Thus, if $p(\betap)$ specifies
that $\betap$ is a Gaussian AR(1) process, then sampling the complete
conditional for $\betap$ is equivalent to sampling $(\betap | \{z_t\}_{t=1}^T)$
from the DLM
\[
\begin{cases}
  z_t = \psi_t + \nu_t, & \nu_t \sim N(0, 1/\omega_t) \\
  \psi_t = \vx_t' \bbeta_t \\
  \bbeta_t = \bmu + \bPhi (\bbeta_{t-1} - \bmu) + \vep_t, & \vep_t \sim N(0, \bW).
\end{cases}
\]

Collecting the complete conditional for $\betap$ and for $\oomega$ leads to
posterior simulation by Gibbs sampling: draw $(\betap|\oomega, \yy)$ using the
FFBS algorithm and draw $(\oomega | \betap, \yy)$ by taking independent samples
of $\omega_t \sim \PG(b_t, \psi_t)$ for $t=1, \ldots, T$.
\citet{polson-etal-2013} describe how to sample from $\PG$ distributions and
implement this sampler in the \texttt{R} package \texttt{BayesLogit}
\citep{bayeslogit-2013}.  Sampling any hyperparameters, like the autocorrelation
coefficient of the AR(1) process or the innovation variance, follows using
standard conjugate or MCMC techniques.

\begin{example}
  Suppose that one observes binomial outcomes $y_t \sim \text{Binom}(n_t, q_t)$
  for $t=1, \ldots, T$.  Letting $\psi_t$ be the log-odds, the data generating
  distribution is
  \[
  p(y_t | \psi_t) = c(y_t) \frac{\exp(\psi_t)^{y_t}}{(1+\exp({\psi_t}))^{n_t}}.
  \]
  Thus the complete conditional $(\betap | y, \omega)$ may be simulated by using
  forward filter backward sampling with pseudo-data $z_t = \kappa_t / \omega_t$
  where $\kappa_t = y_t - n_t / 2$.
\end{example}

\begin{example}
  Suppose that one observes counts according to $y_t \sim \NB(d, q_t)$ for
  $t=1, \ldots, T$, where $d$ is the number of failures before observing $y_t$
  successes, also interpreted as a positive real-valued dispersion coefficient,
  and $q_t$ is the probability of observing a success.  Letting $\psi_t$ be the
  log-odds, the data generating distribution is
  \[
  p(y_t | \psi_t) = c(y_t, d) \frac{\exp({\psi_t})^{y_t}}{(1+\exp({\psi_t}))^{y_t+d}}.
  \]
  In negative binomial regression, it is common to model the log-mean, $\lambda_t
  = \psi_t + \log(d) = \vx_t' \bbeta_t$, instead of the log-odds.  This requires only
  a slight modification.  Following the work above, the complete conditional
  $(\omega_t | \bbeta_t, d)$ is $\PG(b_t, \psi_t)$ where $b_t = y_t + d$.  However,
  the DLM used to estimate $\betap$ is now
  \[
  \begin{cases}
    z_t = \lambda_t + \nu_t, & \nu_t \sim N(0, 1/\omega_t) \\
    \lambda_t = \vx_t' \bbeta_t \\
    \bbeta_t = \bmu + \bPhi (\bbeta_{t-1} - \bmu) + \vep_t, & \vep_t \sim N(0, \bW)
  \end{cases}
  \]
  where $z_t = \kappa_t / \omega_t + \log(d)$ and $\kappa_t = (y_t - d) / 2$.
\end{example}



\begin{figure}
  \begin{center}
    \includegraphics[width=5.5in]{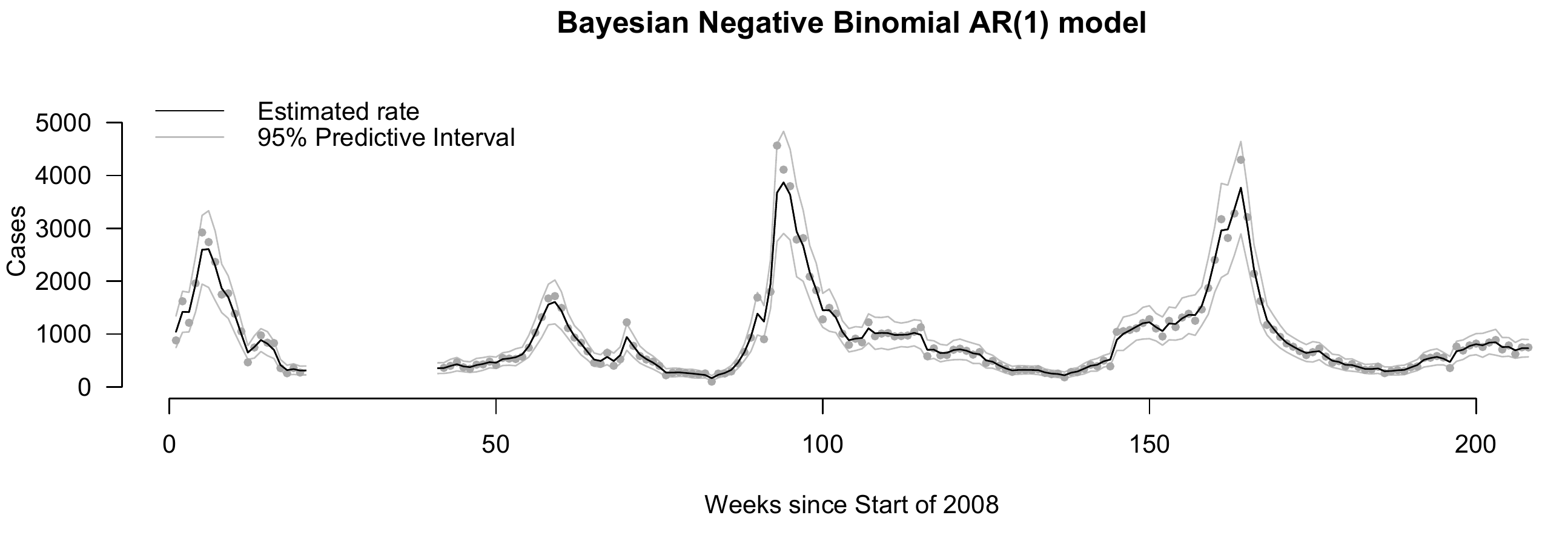}
    \caption{\label{fig:nb-ar1} Incidence of influenza-like illness in Texas,
      2008--11, together with the estimated mean from the negative-binomial AR(1)
      model.  The blanks in weeks 21-41 correspond to missing data.  The grey lines
      depict the upper and lower bounds of a 95$\%$ predictive interval.}
  \end{center}
\end{figure}

As an initial illustration, we fit a negative-binomial AR(1) model to four years
(2008--11) of weekly data on flu incidence in Texas, collected from the Texas
Department of State Health Services.  Let $y_t$ denote the number of reported
cases of influenza-like illness in week $t$.  We assume that these counts follow
a negative-binomial model, which will allow over-dispersion relative to the
Poisson.
Figure \ref{fig:nb-ar1} shows the results of the fit. For simplicity, we assumed
an improper uniform prior on the dispersion parameter $d$, and fixed $\phi$ and
$\sigma^2$ to $0.98$ and 1, respectively, but it is straightforward to place
hyper-priors upon each parameter, and to sample them in a hierarchical fashion.
It is also straightforward to incorporate fixed effects in the form of
regressors.


\section{Comparison}

\subsection{Ease of implementation}

The new data-augmentation approach outlined here may be evaluated along two
axes: ease of use (measured in the fixed cost of time-to-implementation by the
design of a bespoke statistical model), and pure efficiency (measured by runtime
of the algorithm).  Our numerical comparisons show that for all the \Polya-Gamma
technique has excellent effective sampling rates with its strongest competition
coming from the data augmentation approaches of \cite{fussl-etal-2013} and
\cite{fruhwirth-schnatter-etal-2009}.

But pure efficiency is not the only consideration, and we will preface these
results with a few brief comments about the structure of the various algorithms,
especially regarding their ease of use.  Both \cite{fussl-etal-2013} and
\cite{fruhwirth-schnatter-etal-2009} rely on mixture representations of the
data-generating process that are reminiscent of \cite{albert-chib-1993}.  The
difference from the probit model is that, in the logit and negative-binomial
cases, directly mimicking the Albert/Chib missing-data mechanism does not lead
to a conditionally Gaussian representation.  Rather, it leads to an exotic
distribution that must be approximated by a discrete mixture of normal random
variables.  Crucially, the parameters of this exotic distribution depend on the
data.  Therefore, one cannot simply consider a single troublesome distribution,
but rather an entire family of such distributions, and therefore a corresponding
family of approximations.

This family, moreover, is uncountably large, and cannot be captured by a finite
collection of approximations.  One must therefore find a way to interpolate
between approximations to cover the entire parametric space.  That is, in order
for the overall scheme to be practical, one must resort to approximations of
approximations.

The papers by \cite{fussl-etal-2013} and \cite{fruhwirth-schnatter-etal-2009}
show how to stitch these two approximations together to yield an approximate
sampling method for any model in the class.  In both cases, the set of discrete
mixture-of-normals approximations, along with the needed interpolation scheme,
may be precomputed; and it is these extensive prior computations that account
for the method's success on the pure-efficiency axis.  We have not included the
time required by such computations in our comparisons, but this upfront cost,
and the inexactness of the method, must be borne in mind by practitioners.

The \Polya-Gamma approach, meanwhile, is both exact and simple.  It depends only
upon Equation (\ref{eqn:pg-cancellation}), which directly turns an
intractable likelihood into a familiar Gaussian form.  This holds not just for
the AR(1) prior presented previously, but for any linear evolution equation,
such as a dynamic factor model.  Since the requisite Gaussian machinery is well
established, the only challenge is sampling the conditional distribution of the
latent variables.  This, too, requires working with a distribution that will
seem exotic to most Bayesian practitioners, at least initially.  But this cost
has already been borne: the only step here is to simulate \Polya-Gamma random
variates, which can be done using the \texttt{BayesLogit} R Package.  (The C
code used by the \texttt{BayesLogit} package is freely available, so this
applies equally to any scripting language that can make calls to external C
routines.)  Thus, the only novel aspect of implementing the \Polya-Gamma data
augmentation approach is to insert one extra module---effectively, one extra
line of code---into an existing sampling scheme.

It is also worth considering that the Polya-Gamma class is a relatively new
distributional family.  Further research will likely bring a more efficient
sampler that will diminish any gaps in performance.  These gains will be
realized simply by plugging in the new sampling module into existing code.  We
would argue that this blend of simplicity, modularity, and efficiency is ideal
for practitioners.

\subsection{Efficiency}

As reflected in Section \ref{sec:previous-efforts}, there are many alternative
techniques for sampling posterior distributions that arise from modeling
discrete or countable time-varying data.  Here we compare our data augmentation
approach to the best alternatives: \cite{fussl-etal-2013} and
\cite{fruhwirth-schnatter-etal-2009} for binomial logistic regression and
negative binomial regression using data augmentation and \cite{migon-etal-2013}
for the same models using a Metropolis-Hastings based approach.

Since Markov Chains generate correlated samples we compare methods by measuring
how fast each procedure produces \emph{nearly} independent samples, that is we
measure the effective sampling rate (ESR).  To that end, we employ the effective
sample size (ESS), which approximates the number of ``independent'' draws
produced by a sample of $M$ correlated draws.  One may view the ESS as an
estimate of the number of samples produced after having thinned the $M$
correlated samples so that remaining draws appear independent.  From
\citet{holmes-held-2006}, the effective sample size is
\[
ESS_{it} = M / \Big( 1 + 2 \sum_{k=1}^\ell \rho_k(\beta_{it}) \Big)
\]
where $\rho_k(\beta_{it})$ is the $k$th lagged autocorrelation of the chain
corresponding to the $i$th component of $\bbeta_t$.  The effective
sampling rate is the ratio of the effective sample size to the time taken to
generate the post-burn-in samples; thus, it measures the rate at which the
Markov Chain produces independent draws after initialization and burn-in.

To mitigate MCMC sample variation, 10 batches of 12,000 samples are taken and
the last 10,000 draws are recorded.  For batch $m$, we compute the
component-wise effective sample size $ESS_{it}^{m}$ corresponding to the
univariate chain for $\beta_{it}$.  Taking the mean over batches produces the
average, component-wise effective sample size $\overline{ESS}_{it}$ and, after
normalizing each batch by time, the average, component-wise effective sampling
rate $\overline{ESR}_{it}$.  Following,
\citet{fruhwirth-schnatter-fruhwirth-2010}, the primary metric of comparison is
the median effective sampling rate,
\[
\text{med} \; \Big\{ \overline{ESR}_{it} : i=1, \ldots, P; \;  t=1, \ldots, T \Big\}.
\]

We consider synthetic data sets with a variety of characteristics.
For dynamic binomial logistic regression, we consider log odds of the form
$\alpha + \vx_t' \bbeta_t$ where $\alpha$ is a scalar and $\{\bbeta_t\}_{t=1}^T$
is a 2-dimensional AR(1) process with autocorrelation coefficient $\bPhi = 0.95
I_2$.  Four different synthetic data sets are constructed, allowing the
covariates $\{\vx_t\}_{t=1}^T$ to have lots or little of correlation and letting
the responses $y_t$ arise from $\text{Binom}(n, q_t), t=1, \ldots, T$ with
either $n=1$ or $n=20$ trials.  The setup is almost identical for dynamic
negative binomial regression except that we model the log-mean as $\alpha +
\vx_t' \bbeta_t$ and consider responses with $\alpha = \log(10)$ or $\alpha =
\log(100)$ corresponding to average count sizes of roughly 10 or 100.  Further
details may be found in the supplementary material.

Some caution is warranted when comparing methods as the effective sampling rate
is sensitive to a procedure's implementation and the hardware on which it is
run.  (Supplementary files have been provided so that users may examine the
performance of these methods on their own.)  The routines are written primarily
in R.  We use code from \citet{binomlogit-2012} and
\citet{fruhwirth-schnatter-book-2007} for the discrete mixture of normals
methods.  All benchmarks are carried out on an Ubuntu machine with Intel Core
i5-3570 3.4GHz processor and 8GB of RAM.  Some computations were burdensome in
R, and hence we wrote wrappers to C to speed up the MCMC simulations.  In
particular, both data augmentation methods implement forward filtering and
backward sampling using a C wrapper.  The \Polya-Gamma technique calls C code to
sample random variates using version 0.2-4 of the \texttt{BayesLogit} package
\citep{bayeslogit-2013}.  The conjugate updating and backwards sampling of
\citet{migon-etal-2013} is done in C.  Having a C wrapper to forward filter and
backwards sample is particularly important, as our C implementation is much
faster than the corresponding R code.  Were we to use the slower R version, our
results would favor the \Polya-Gamma method, as it has better effective sample
sizes and would spend less time, proportionally, sampling the latent random
variables.

\citet{polson-etal-2013} outline the expected performance of the \Polya-Gamma
data augmentation technique, which depends heavily on how quickly one can
generate \Polya-Gamma random variates.  In their original algorithm, sampling
\Polya-Gamma random variates from $\PG(b,\psi)$ is fast when $b$ is a small
integer, but slow when $b$ is large. \citet{windle-thesis-2013} provides an
improved method for sampling $\PG(b,\psi)$; however sampling large $b$ is still
slower than sampling small $b$.  These differences in computational cost are
important, as one must sample $\omega_t \sim \PG(b_t, \psi_t), t=1,\ldots,T$,
for each MCMC iteration under \Polya-Gamma data augmentation.  In binomial
logistic regression, $b_t = n_t$ where $n_t$ is the number of trials at each
response $y_t$.  Hence, when there are few trials, as is usually the case, the
PG method will do well.  For negative binomial regression $b_t = y_t + d_t$
where $y_t$ is the response and $d_t$ is the dispersion.  Thus, larger average
counts sizes will lead to longer MCMC simulations.

In general, we find these principles to hold.  The \Polya-Gamma data
augmentation technique performs well for dynamic binomial logistic regression
when the number of trials of the response is small, showing a roughly 25\%
better ESR for binary logistic regression than \citet{fussl-etal-2013}; however,
\citet{fussl-etal-2013} does slightly better when the number of trials is large.
Similarly, the \Polya-Gamma technique outpaces
\citet{fruhwirth-schnatter-etal-2009} in negative binomial regression when the
average number of counts is small, but loses when the average number of counts
is large.  The Metropolis-Hastings approach of \citet{migon-etal-2013} does the
worst in all of the tests.  Part of its poor performance is due to a non-linear
solve that must be made $T$ times when forward filtering.  We did not attempt to
optimize the performance of this non-linear solver, and hence some improvement
may be possible, though the disparity in ESRs suggests that any improvement
would not be enough to compete with either data augmentation approach.  As a
check upon \citet{migon-etal-2013}, we also implemented a Metropolis-Hastings
sampler that draws blocks of disturbances using Laplace approximations.  This
fared worse still.

Of note, the \Polya-Gamma method almost always has superior effective sample
sizes.  Hence faster \Polya-Gamma samplers could push the \Polya-Gamma data
augmentation technique to the top for all of the models considered.  Table
\ref{tab:benchmark-summary} provides a summary of the benchmarks; details may be
found in the supplementary material.

\begin{table}

  \begin{center}
    \small
    \begin{tabular}{l l c c}
      \multicolumn{4}{c}{Dynamic Binom.\ Logistic Reg.} \\
      \hline
      & & Est.\ & ESR \\
      n & f & AR & PG/dRUM \\
      \hline
      $1$ & low & no & 1.26 \\
      $1$ & low & yes & 1.30 \\

      $1$ & high & no & 1.23 \\
      $1$ & high & yes & 1.23 \\

      $20$ & low & no & 0.91 \\
      $20$ & low & yes & 0.98 \\

      $20$ & high & no & 0.83 \\
      $20$ & high & yes & 0.98
    \end{tabular}
    \hspace{12pt}
    \begin{tabular}{l l c c}
      \multicolumn{4}{c}{Dynamic Neg.\ Binomial Reg.} \\
      \hline
      & & Est.\ & ESR \\
      $\mu$ & f & AR & PG/FS \\
      \hline
      $10$ & low & no & 1.03 \\
      $10$ & low & yes & 1.86 \\

      $10$ & high & no & 1.15 \\
      $10$ & high & yes & 1.06 \\

      $100$ & low & no & 0.76 \\
      $100$ & low & yes & 0.82 \\

      $100$ & high & no & 0.76 \\
      $100$ & high & yes & 0.78 \\

    \end{tabular}
  \end{center}

  \begin{center}
    \caption{\label{tab:benchmark-summary} A summary of the benchmarking
      results. ESR PG/dRUM and ESR PG/FS report the ratio of the median
      effective sampling rate for the PG method compared to the best
      alternative, which in both cases is the discrete mixture of normal
      approaches \citep{fussl-etal-2013, fruhwirth-schnatter-etal-2009}.  A
      higher ratio corresponds to the PG method doing better.  $n$ corresponds
      to the number of trials for the binomial response while $\mu$ corresponds
      to the approximate mean for the negative binomial response.  $f$
      determines whether there is a low or high amount of correlation between
      the covariates.  Est.\ AR indicates whether the parameters of the AR
      process were estimated or not.  }

  \end{center}

\end{table}

\section{Conclusion}

The \Polya-Gamma data augmentation approach to dynamic models of counts or
categories is elegant, efficient, and leads to familiar complete conditionals in
the quantity of interest, making it easy to implement and customize.  Thus, it
is an excellent choice for almost any modeling scenario.  For those concerned
most with computational efficiency, there is one caveat to this conclusion: the
approach of \cite{fruhwirth-schnatter-etal-2009} can lead to faster raw
computations, at the cost of the implementational challenges already described.

\bibliography{bayeslogit}{}
\bibliographystyle{abbrvnat}

\end{document}